\documentclass[12pt,onecolumn,prc,showpacs,preprintnumber,amsmath,
floatfix,amssymb]{revtex4}
\usepackage{epsfig}
\usepackage{graphicx}
\usepackage{dcolumn}
\usepackage{bm}
\def\var{\mbox{\boldmath $\varepsilon$}}
\def\p{\mbox{\boldmath $p$}}
\def\q{\mbox{\boldmath $q$}}
\def\k{\mbox{\boldmath $k$}}

\allowdisplaybreaks[4]

\begin{document}
\title{Quasi-elastic neutrino charged-current scattering off $^{12}$C: effects 
 of the meson exchange currents and large nucleon axial mass}
\author{A.~V.~Butkevich$^{1}$ and S.~V.~Luchuk$^{1,2}$}
\affiliation{$^{1}$Institute for Nuclear Research,
Russian Academy of Sciences, Moscow 117312, Russia\\
$^{2}$Moscow Institute of Physics and Technology, Dolgoprudny 141701, Russia}
\date{\today}
\begin{abstract}

The quasi-elastic scattering of muon neutrino and electrons 
on a carbon target are analyzed using the relativistic distorted-wave 
impulse approximation (RDWIA). We also evaluate the contribution of the 
two-particle and two-hole meson exchange current ($2p-2h$ MEC) to electroweak 
response functions. The nuclear model dependence of the (anti)neutrino cross 
sections is studied within the RDWIA+MEC approach and RDWIA model with the 
large nucleon axial mass. It is shown that the results for the squared  
momentum transfer distribution $d\sigma/dQ^2$ and for invariant mass of 
the final hadronic system distribution $d\sigma/dW$ obtained within these 
models are substantially different.    

\end{abstract}

\maketitle

\section{Introduction}

One of the important goals of the current~\cite {NOvA1, T2K1} and upcoming~
\cite {DUNE, HK} accelerator-based neutrino experiments is the determination 
of the neutrino masses ordering. The question is weather we have 
two ``light'' and one ``heavy'' neutrino ( the so-called normal mass hierarchy)
 or two ``heavy'' and one ``light'' neutrino (the inverted hierarchy). When 
neutrino propagate through a medium the oscillation physics is modified by the 
so-called matter effect~\cite{Wol, MS}. Matter effects depend on the ordering 
of the neutrino mass eigenstates and allow one to probe the mass hierarchy in  
different ways. Thanks to the matter effects in the Sun, we know 
that $\nu_1$ is lighter than $\nu_2$, where $(\nu_1, \nu_2, \nu_3)$ are 
neutrino with well-defined masses. For $\nu_1(\nu_2)$ and $\nu_3$ sector, 
matter effects in the Earth's crust are significant (about 30\%) for neutrino 
energy $\varepsilon_{\nu} \sim 1\div 5$ GeV and propagation distance 
$L \sim 10^3$ km. 

In this energy regime the dominant contribution to the neutrino-nucleus cross 
section comes from the charged-current (CC) quasielastic (CCQE) 
reactions, two-body meson exchange current (MEC), and resonance production 
processes. To evaluate the neutrino mass-square difference in muon neutrino 
oscillation experiments, the probabilities of $\nu_{\mu}$ disappearance and 
$\nu_e$ appearance versus neutrino energy are measured.
At neutrino energy $\varepsilon_{\nu} \geq 2$ GeV the contribution of the CCQE 
scattering is less than 40\% and therefore the incoming neutrino energy is 
estimated applying the calorimetric energy reconstruction method, 
already actively used in experiments. 

Conservation of total energy in CC neutrino interactions implies
$\varepsilon_{\nu}=\varepsilon_f + \varepsilon_h$, where $\var_{f}$ and 
$\var_h$ are lepton and hadronic energies, respectively.
Thus, the total hadronic energy deposit is the necessary information for 
calorimetric method. Muon energy is reconstructed from the measured path 
length in the detector. Hadronic energy is obtained from calorimetry by  
first summing all the visible energy not associated with the muon. However, it 
is impossible to measure energies of all hadrons, notable energy deposits by 
neutrons are always hard to measure. A model dependent fit obtained from 
simulation is used to relate the summed visible energy to the estimated total 
hadronic energy. This procedure is not free from systematic uncertainties 
affecting the determination of the incident neutrino energy. For instance, the 
estimated muon and hadronic energy resolution are 3.5\% and 25\%, respectively,
 giving an overall energy resolution for selected $\nu_{\mu}$-CC events of
 about 7\% for fully active and fine-grained NOvA detectors~\cite{NOvA2}. 

In addition to its role in reconstruction of the incident neutrino energy, the 
hadronic energy plays an essential role in studying CCQE interactions. 
Note, as the quasielastic interaction is a two-particle scattering process,
it forms a CCQE interaction sample and energy of the incoming neutrino can be 
derived from lepton kinematics alone. The measurement of muon momentum and 
angle allows one to estimate neutrino energy $\varepsilon^{QE}_{\nu}$ and the 
squared four-momentum transfer $Q^2_{QE}$, assuming the target nucleon at rest. 
This reconstruction method (kinematic method) works well if the true nature of 
events were indeed a CCQE process. 
Therefore accuracy of the kinematic neutrino energy reconstruction method 
depends on the purity of the CCQE sample, and therefore measurements of the 
differential $d\sigma/dQ^2_{QE}$ and total $\sigma(\varepsilon^{QE}_{\nu})$ 
cross sections are model dependent. Note, the calorimetric method can also be 
applied to the CCQE events. Modern neutrino experiments are investigating the 
axial-vector current contribution to the quasielastic neutrino scattering on 
nuclei. For estimation of neutrino energy, the kinematic method is applied. 
Using the dipole parametrization of the axial form factor and the values of 
$\var^{QE}_{\nu}$ and $Q^2_{QE}$, these experiments have extracted within the 
relativistic Fermi gas model (RFGM)~\cite{Moniz} the values of 
$M_A\approx 1.2 \div 1.4 $ GeV, that are systematically higher than 
$M_A\approx 1$ GeV obtained from deuterium target.  

These results have stimulated many theoretical studies trying to explain 
the apparent discrepancies between data and theoretical predictions. A detailed
 review of the experimental results and theoretical framework of 
neutrino-nucleus CCQE-like interaction can be found elsewhere (see for instance
~\cite{Garvey, Katori} and references therein). Based on the results 
from different groups~\cite{Martini1, Martini2, Nieves1, Nieves2, Pace, 
Megias1, Simo, Megias2} it is shown that CCQE-like data are really a 
combination of genuine QE and $np-nh$ contributions. The inclusion of 
two-particle and two-hole ($2p-2h$) meson exchange current (MEC) contributions, 
has allowed one to explain experimental results without modification of the 
axial mass, (i.e., with $M_A\approx 1$ GeV). The MEC 
effects play an important role in the ``dip'' region between the QE and 
$\Delta$ peaks, where the energy of the hadronic final system produced in the 
two-body MEC processes is larger than in the CCQE interaction. That is why, 
there is a growing interest in utilizing hadron information to study MEC 
contributions. On the other hand, with detector that can directly measure at 
least a part of the hadronic energy, the $2p-2h$ contribution to the CCQE 
events sample can be reduced. In this case the MEC contributions are treated as 
background, and one would expect CCQE sample to provide the cross 
sections that more or less agree with the RFGM predictions with $M_A\approx 1$ 
GeV (the so-called golden scenario). 

In this work we perform a joint calculation of the CCQE and $2p-2h$ 
contributions to lepton scattering cross sections on carbon, using the 
relativistic distorted-wave impulse approximation (RDWIA) with $M_A=1.03$ GeV 
 for quasielastic responses and the $2p-2h$ meson exchange currents response 
functions in the electroweak sector (RDWIA+MEC prediction). We also calculate 
(anti)neutrino cross sections within the RDWIA approach with $M_A=1.35$ GeV. 
The RDWIA, which takes into account the nuclear shell structure and final state 
interaction (FSI) effects, was developed for description of QE electron-nucleus
 scattering and was successfully tested against the 
data~\cite{Kelly1, Fissum, Kelly2}. This approach was also applied to 
neutrino-nucleus interactions to calculate the genuine QE cross sections
~\cite{Meucci, Maieron, BAV1, BAV2, BAV3, BAV4}. In our approach~
\cite{BAV1, BAV2} the effects of the short-range nucleon-nucleon ($NN$)- 
correlations in the nuclei ground state are estimated.   
 
To evaluate the MEC response we use simple parameterizations of the exact
 MEC calculations of the electroweak response as functions of the momentum and 
energy transfer. These calculations were performed within RFGM. The functional 
forms employed for the parameterizations of the transverse electromagnetic 
vector response, and for axial and vector components of the weak response were 
detailed in~\cite{Megias1, Megias2}. These parameterizations have been 
validated by describing the full set data of inclusive cross section of 
electron scattering on carbon~\cite{Megias3} and data from the neutrino 
experiments~\cite{Megias2}. The results show good agreement with experimental 
data over wide range of energy transfer. 

The aim of this work is twofold. First, we test the RDWIA+MEC approach with 
${}^{12}$C$(e,e')$ scattering data for different kinematic situations. The 
accordance between this model predictions and data in the vector sector gives 
us confidence in the extension of this phenomenological approach and its 
validity, when applied to calculation of the CCQE-like cross sections of the 
(anti)neutrino scattering on carbon. Second, 
we compare the neutrino cross sections calculated in the RDWIA+MEC and RDWIA 
(with $M_A=1.35$ GeV) approaches to study the effects due to the MEC 
contributions and large nucleon axial mass. This issue is very important for 
neutrino oscillation experiments, provided that the two effects (whether one 
changes the transverse response or axial form factor) have very different 
consequences on the energy dependence of the CCQE cross section and the 
determination of $\varepsilon_{\nu}$.     

The paper is organized as follows. In Sec. II we briefly introduce the 
formalism needed for studying quasielastic lepton scattering off nuclei with 
the $2p-2h$ MEC contributions, and describe the RDWIA model and our MEC 
calculations. The results are presented and discussed in Sec. III. Our 
conclusions are summarized in Sec. IV. 

\section{Formalism of quasielastic scattering, RDWIA, and $2p-2h$ MEC 
responses}

We consider electron and neutrino charged-current QE inclusive
\begin{equation}\label{qe:incl}
\l(k_i) + A(p_A)  \rightarrow \l'(k_f) + X                      
\end{equation}
scattering off nuclei in the one-photon (W-boson) exchange approximation. Here 
$l$ labels the incident lepton [electron or muon (anti)neutrino], and
$l^{\prime}$ represents the scattered lepton (electron or muon),
$k_i=(\varepsilon_i,\k_i)$ and $k_f=(\varepsilon_f,\k_f)$ are the initial and 
final lepton momenta, $p_A=(\varepsilon_A,\p_A)$ is 
the initial target momenta, $q=(\omega,\q)$ is the momentum transfer carried by 
the virtual photon (W-boson), and $Q^2=-q^2=\q^2-\omega^2$ is the photon 
(W-boson) virtuality. 

\subsection{Quasielastic lepton-nucleus cross sections}

In the inclusive reactions (\ref{qe:incl}) only the outgoing lepton is
detected and the differential cross sections can be written as
\begin{subequations}
\begin{align}
\frac{d^3\sigma^{el}}{d\varepsilon_f d\Omega_f} &=
\frac{\varepsilon_f}{\varepsilon_i}
 \frac{\alpha^2}{Q^4} L_{\mu \nu}^{(el)}{W}^{\mu \nu (el)},
\\                                                                       
\frac{d^3\sigma^{cc}}{d\varepsilon_f d\Omega_f } &=
\frac{1}{(2\pi)^2}\frac{\vert\k_f\vert}
{\varepsilon_i} \frac{G^2\cos^2\theta_c}{2} L_{\mu \nu}^{(cc)}
{W}^{\mu \nu (cc)},
\end{align}
\end{subequations}
where $\Omega_f=(\theta,\phi)$ is the solid angle for the lepton momentum, 
$\alpha\simeq 1/137$ is the fine-structure constant, 
$G \simeq$ 1.16639 $\times 10^{-11}$ MeV$^{-2}$ is
the Fermi constant, $\theta_C$ is the Cabbibo angle
($\cos \theta_C \approx$ 0.9749), $L_{\mu \nu}$ is the lepton tensor,
 $W^{\mu \nu (el)}$ and $W^{\mu \nu (cc)}$ are correspondingly the 
electromagnetic and weak CC nuclear tensors. In terms of response functions the 
cross sections reduce to
\begin{subequations}
\begin{align}
\frac{d^3\sigma^{el}}{d\varepsilon_f d\Omega_f} &=
\sigma_M\big(V_LR^{(el)}_L + V_TR^{(el)}_T\big),
\\                                                                       
\frac{d^3\sigma^{cc}}{d\varepsilon_f d\Omega_f} &=
\frac{G^2\cos^2\theta_c}{(2\pi)^2} \varepsilon_f
\vert \k_f \vert\big ( v_0R_0 + v_TR_T
+ v_{zz}R_{zz} -v_{0z}R_{0z}- hv_{xy}R_{xy}\big),
\end{align}
\end{subequations}
where 
\begin{equation}
\sigma_M = \frac{\alpha^2\cos^2 \theta/2}{4\varepsilon^2_i\sin^4 \theta/2} 
\end{equation}
is the Mott cross section. The electron $V_k$ and neutrino $v_k$ coupling 
coefficients, whose expressions are given in~\cite{BAV1} are kinematic 
factors depending on the lepton's kinematics. The response functions are given 
in terms of components of the hadronic tensors
\begin{subequations}
\begin{align}
R^{(el)}_L &=W^{00 (el)},
\\
R^{(el)}_T &=W^{xx (el)}+W^{yy (el)},                                  
\\
R_0 & = W^{00(cc)},\\
R_T & = W^{xx(cc)} + W^{yy(cc)},\\
R_{0z}&  = W^{0z(cc)} + W^{z0(cc)},\\
R_{zz} & = W^{zz(cc)}, \\                                             
R_{xy} & =  i\left(W^{xy(cc)}-W^{yx(cc)}\right), 
\end{align}
\label{R}
\end{subequations}
and depend on the variables ($Q^2, \omega$) or ($|q|,\omega$). They describe 
the electromagnetic and weak properties of the hadronic system.  

All the nuclear structure information and final state interaction effects 
(FSI) are contained in the electromagnetic or weak CC nuclear tensors. They 
are given by the bilinear products of the transition matrix 
elements of the nuclear electromagnetic or CC operator $J^{(el)(cc)}_{\mu}$ 
between the initial nucleus state $|A\rangle$ and the final state 
$|X\rangle$ as 
\begin{eqnarray}
W_{\mu \nu } &=& \sum_f \langle X\vert                           
J^{(el)(CC)}_{\mu}\vert A\rangle \langle A\vert
J^{(el)(CC)\dagger}_{\nu}\vert X\rangle,              
\label{W}
\end{eqnarray}
where the sum is taken over undetected states $X$. This equation is very 
general and includes all possible final states. Thus, the hadron tensors can be
expanded as the sum of $1p-1h$ and $2p-2h$, plus additional channels: 
\begin{eqnarray}
W^{\mu \nu } &=& W^{\mu \nu}_{1p1h} + W^{\mu \nu}_{2p2h} + \cdots        
\label{W_12}
\end{eqnarray}
In the impulse approximation (IA) the $1p-1h$ channel gives the well-known 
CCQE response functions and the $2p-2h$ hadronic tensor determines the
 $2p-2h$ MEC response functions. Thus, the functions $R_i$~(\ref{R}) can be 
written as a sum of the CCQE ($R_{i,QE}$) and MEC ($R_{i,MEC}$) response functions
\begin{eqnarray}
R_i &=& R_{i,QE} + R_{i,MEC}         
\label{R_12}
\end{eqnarray}

\subsection{RDWIA model}

We describe genuine CCQE neutrino-nuclear scattering in the impulse 
approximation (IA), assuming that the incoming neutrino interacts with only 
one nucleon, which is subsequently emitted, while the remaining (A-1) nucleons 
in the target are spectators. The nuclear current is written as the sum of 
single-nucleon currents.

For electron scattering, we use the CC2 electromagnetic vertex
function for a free nucleon~\cite{deFor}
\begin{equation}
\Gamma^{\mu}_V = F^{(el)}_V(Q^2)\gamma^{\mu} + {i}\sigma^{\mu \nu}\frac{q_{\nu}}
{2m}F^{(el)}_M(Q^2),                                                          
\end{equation}
where $\sigma^{\mu \nu}=i[\gamma^{\mu},\gamma^{\nu}]/2$, $F^{(el)}_V$ and
$F^{(el)}_M$ are the Dirac and Pauli nucleon form factors. 
The single-nucleon charged current has $V{-}A$ structure $J^{\mu(cc)} = 
J^{\mu}_V + J^{\mu}_A$. For a free-nucleon vertex function 
$\Gamma^{\mu(cc)} = \Gamma^{\mu}_V + \Gamma^{\mu}_A$ we use the CC2 vector 
current vertex function
\begin{equation}
\Gamma^{\mu}_V = F_V(Q^2)\gamma^{\mu} + {i}\sigma^{\mu \nu}\frac{q_{\nu}}
{2m}F_M(Q^2)                                                          
\end{equation}
and the axial current vertex function
\begin{equation}
\Gamma^{\mu}_A = F_A(Q^2)\gamma^{\mu}\gamma_5 + F_P(Q^2)q^{\mu}\gamma_5. 
\end{equation}
The weak vector form factors $F_V$ and $F_M$ are related to the corresponding 
electromagnetic form factors $F^{(el)}_V$ and $F^{(el)}_M$ for protons and 
neutrons by the hypothesis of the conserved vector current. We use the 
approximation of Ref.~\cite{MMD} for the Dirac and Pauli nucleon form factors. 
Because the bound nucleons are off-shell we employ the de Forest 
prescription~\cite{deFor} and Coulomb gauge for off-shell vector current 
vertex $\Gamma^{\mu}_V$. 
The vector-axial $F_A$ and pseudoscalar $F_P$ form 
factors are parametrized using a dipole approximation: 
\begin{equation}
F_A(Q^2)=\frac{F_A(0)}{(1+Q^2/M_A^2)^2},\quad                         
F_P(Q^2)=\frac{2m F_A(Q^2)}{m_{\pi}^2+Q^2},
\end{equation}
where $F_A(0)=1.267$, $M_A$ is the axial mass, which controls $Q^2$-dependence 
of $F_A(Q^2)$, and $m_\pi$ is the pion mass.
 
In the RDWIA, the relativistic wave functions of the bound nucleon states are 
calculated in the independent particle shell model as the self-consistent 
solutions of a Dirac equation, derived within a relativistic mean field 
approach, from a Lagrangian containing $\sigma, \omega$, and $\rho$ mesons 
(the $\sigma-\omega$ model)\cite{Serot,Horow}. 
According to the JLab data~\cite{Dutta, Kelly2} the occupancy of the 
independent particle shell-model orbitals of ${}^{12}$C equals on 
average 89\%. In this work, we assume that the missing strength (11\%) can be 
attributed to the short-range $NN$-correlations in the 
ground state, leading to the appearance of the high-momentum and 
high-energy component in the nucleon distribution in the target.  
These estimates of the depletion of hole states are consistent with a direct 
measurement of the spectral function~\cite{Rohe}, which observed approximately 
0.6 protons in a region attributable to a single-nucleon knockout from a 
correlated cluster. 
In the RDWIA, final state interaction effects for the outgoing nucleon are 
taken into account. The distorted-wave function of the knocked out nucleon is 
evaluated as a solution of a Dirac equation containing a phenomenological 
relativistic optical potential. 
The LEA program~\cite{LEA} for the numerical calculation of the 
distorted wave functions with the EDAD1 parametrization~\cite{Cooper} of the 
relativistic optical potential for carbon was used. We calculated the inclusive 
and total cross sections with the EDAD1 relativistic optical potential in 
which only the real part was included.

The cross sections with the FSI effects in the presence of the 
short-range $NN$-correlations were calculated by using the method proposed in
 Ref.~\cite{BAV1} with the nucleon high-momentum distribution from 
Ref.~\cite{Atti} that was renormalized to value of 11\%. In this approach, the 
contribution of the $NN$-correlated pairs is evaluated in impulse 
approximation, i.e., the virtual photon (W-boson) couples to only one member 
of the $NN$-pair. It is a one-body process that leads to the emission of two 
nucleons ($2p-2h$ excitation). 

\subsection{$2p-2h$ excitation}

In order to evaluate the $2p-2h$ hadronic tensor $W^{\mu \nu}_{2p 2h}$, in 
Refs.~\cite{Pace,Simo} the RFGM was chosen to describe the nuclear ground 
state. The short-range $NN$-correlations and FSI effects were not considered in 
this approach. The elementary hadronic tensor is given by bilinear product of 
the matrix elements of the two-body electromagnetic or weak 
(containing vector and axial components) MEC. Only one-pion exchange is 
included. 

The two-body current operator is obtained from the electroweak pion 
production amplitudes for the nucleon~\cite{Her} with coupling a second 
nucleon to the emitted pion. The pion-production amplitudes are derived in the 
non-linear $\sigma$-model for the $\gamma(W)N \to N'\pi $ reaction together 
with electroweak excitation of the $\Delta(1232)$ resonance and its subsequent 
decay into $\pi N$. 
The resulting MEC 
operator can be written as a sum of seagull, pion-in-flight, pion-pole, and 
Delta-pole operators. The $\Delta$-peak is the main contribution to the pion 
production cross section. But inside the nucleus $\Delta$ can also decay 
into one nucleon that re-scatters producing two-nucleon emission without pions. 
Therefore, this decay of $\Delta$ should be considered as a part of the $2p-2h$ 
channel because $\Delta$ emission already includes $2p-2h$ decay inside the 
nucleus. Consequently, there is no unique way of separating the $\Delta$ 
emission from the $2p-2h$ channels. In Refs.~\cite{Pace,Simo}, to separate 
$2p-2h$ channels, the imaginary part of the Delta propagator was subtracted and 
included into phenomenological inelastic contribution to the cross section. 
As a result, the MEC peak is located in the ``dip'' region between the QE and 
Delta peaks, i.e., the invariant mass of the pion-nucleon pair
 $W^2=(q+p_A)^2=m^2 + 2m\omega -Q^2$  
varies in the range $(m_{\pi}+m)\leq W \leq 1.3-1.4$ GeV.

Each one of the four MEC operators can be decomposed as a sum of vector and 
axial-vector currents. In the axial part only the leading contribution to the 
axial-vector vertex proportional to the form-factor $C^V_5$ is included. This 
form-factor is parametrized as~\cite{Her}
\begin{equation}
\label{G5}
G^5_A = \frac{1.2}{(1+Q^2/M^2_{A\Delta})^2}\cdot \frac{1}{1+Q^2/(3M^2_{A\Delta})},
\end{equation}
with $M_{A\Delta}=1.05$ GeV.

\begin{figure*}
  \begin{center}
    \includegraphics[height=16cm,width=16cm]{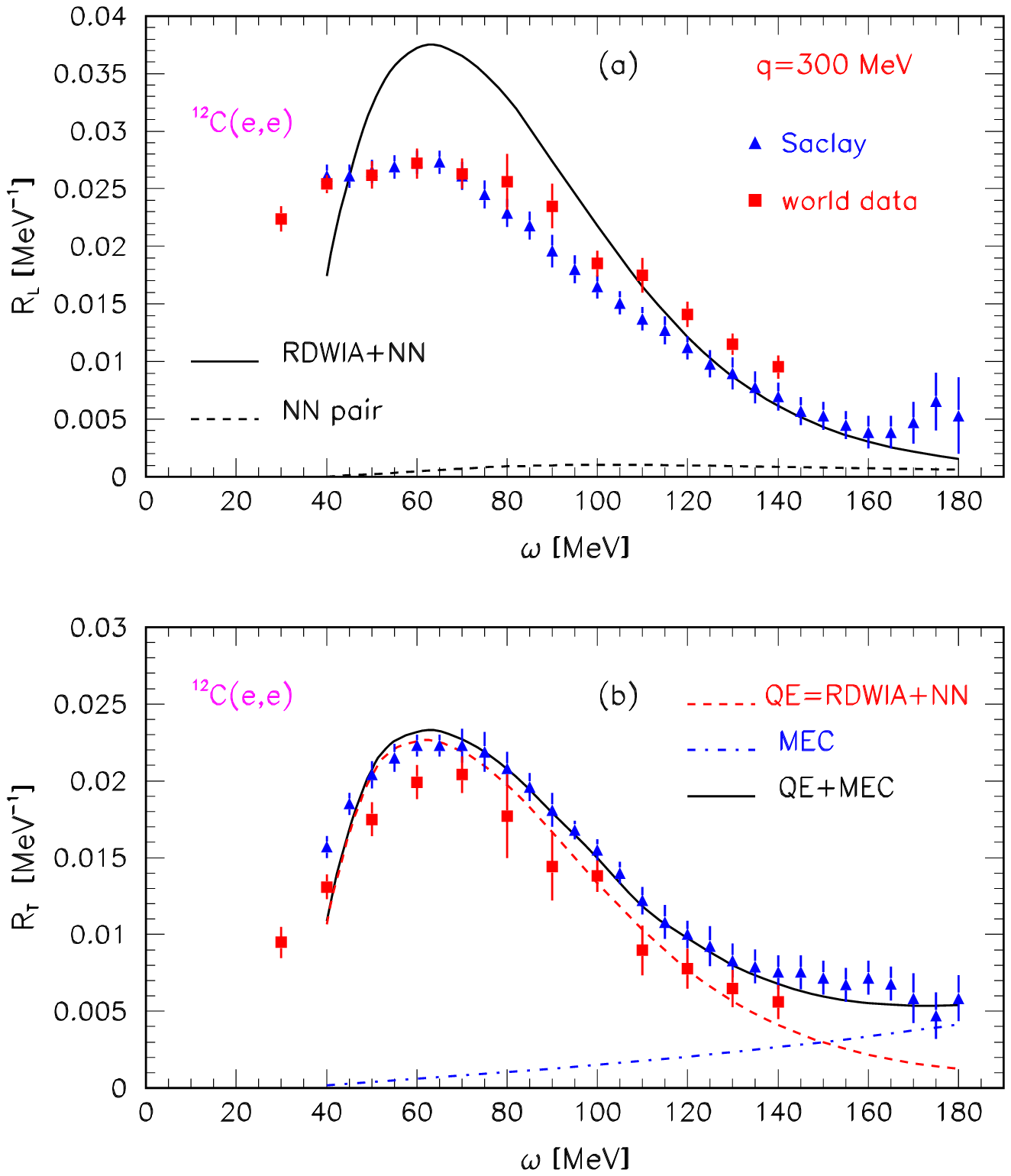}
  \end{center}
  \caption{\label{Fig1}(Color online) Longitudinal (a) and transverse (b) 
response functions at $|\q|=300$ MeV/c versus energy transfer $\omega$ 
 for electron scattering on ${}^{12}C(e,e')$. The solid line is the 
the RDWIA + MEC results, the dashed line is: the contribution from the 
$NN$-correlated pairs in (a) and the contribution from the RDWIA in (b). 
The dash-dotted line in (b) is the $2p-2h$ MEC contributions. The data points 
are from~Refs.\cite{Saclay1,Jour}.} 
\end{figure*}
\begin{figure*}
  \begin{center}
    \includegraphics[height=16cm,width=16cm]{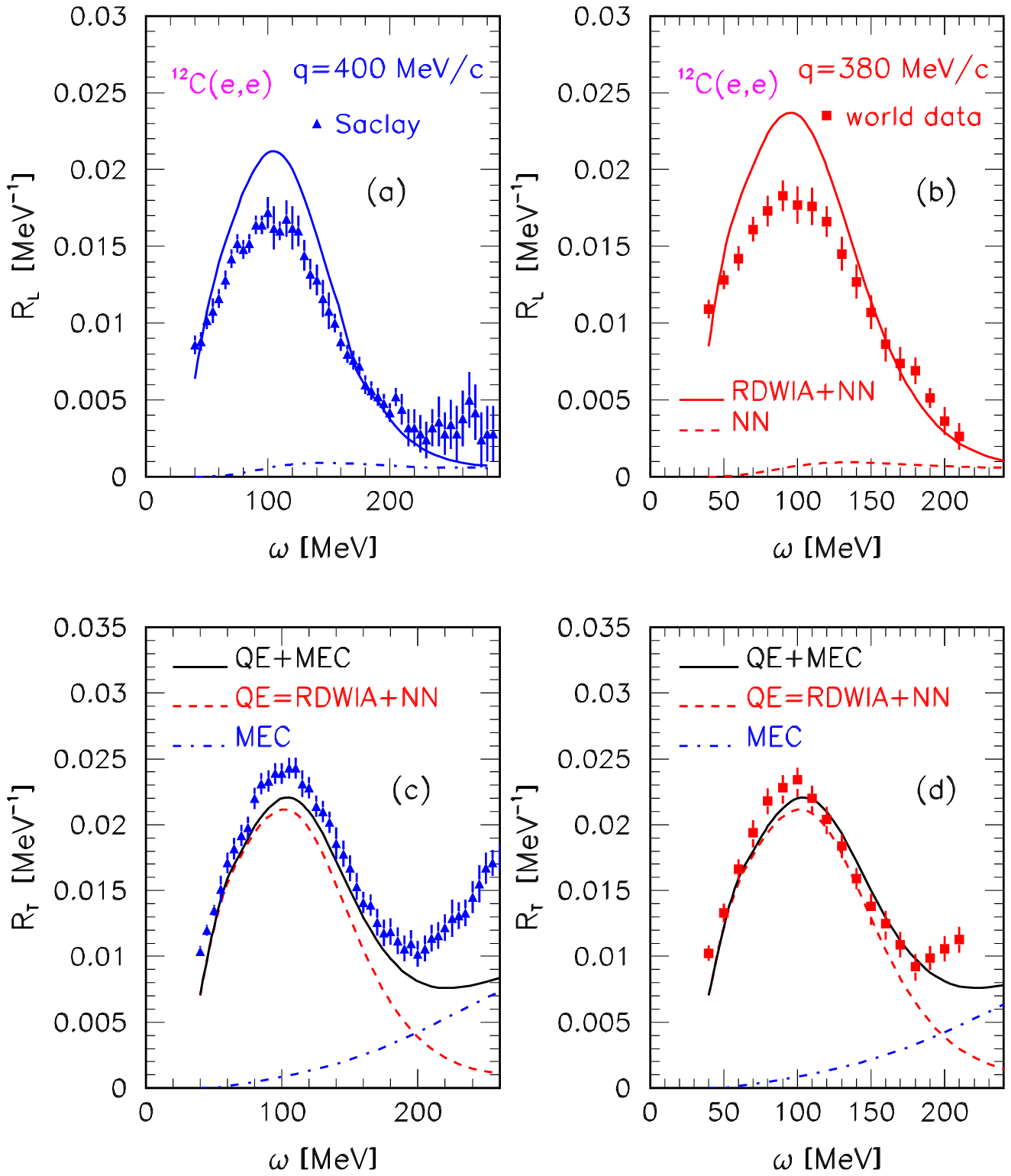}
  \end{center}
  \caption{\label{Fig2}(Color online) Same as in Fig.~\ref{Fig1}, but the  
longitudinal and transverse responses at $|\q|=400$ MeV/c are shown in 
(a), (c) and at $|\q|=380$ MeV/c in (b), (d). As shown in the key the 
data points are from Refs.~\cite{Saclay1,Jour}.}
\end{figure*}

In the present work we evaluate the electroweak MEC response functions 
$R_{i, MEC}$ of lepton scattering on carbon, using accurate parametrizations of 
the exact MEC calculations~\cite{Pace,Simo}. The functional forms employed for 
these parametrizations as functions of $(\omega, |\q|)$  are  valid in the 
range of momentum transfer $|\q|=200 \div 2000$ MeV. The expressions for the 
fitting parameters are described in detail in Refs.~\cite{Megias1, Megias2, 
MeAm}. Results of lepton-nucleus cross sections obtained using these MEC 
parametrizations were successfully tested against the experimental world data
 for ${}^{12}$C~\cite{Megias2, Megias3}. 

\section{Results and analysis}

Before providing reliable predictions for neutrino scattering, any model must be
 validated by confronting it with quasielastic electron scattering data.
To validate the RDWIA+MEC prescription we compare our results for the 
longitudinal and transverse responses, as well as for inclusive $(e,e')$ cross 
sections, with experimental data. A consistent evaluation of these responses 
and cross sections is critical for a proper analysis of neutrino-nucleus 
interaction, as it allows to assess the validity of the RDWIA+MEC approach, 
at least in the vector sector.   

\subsection{Electromagnetic response functions and ${}^{12}$C$(e,e')$ cross 
  sections}

\begin{figure*}
  \begin{center}
    \includegraphics[height=16cm,width=16cm]{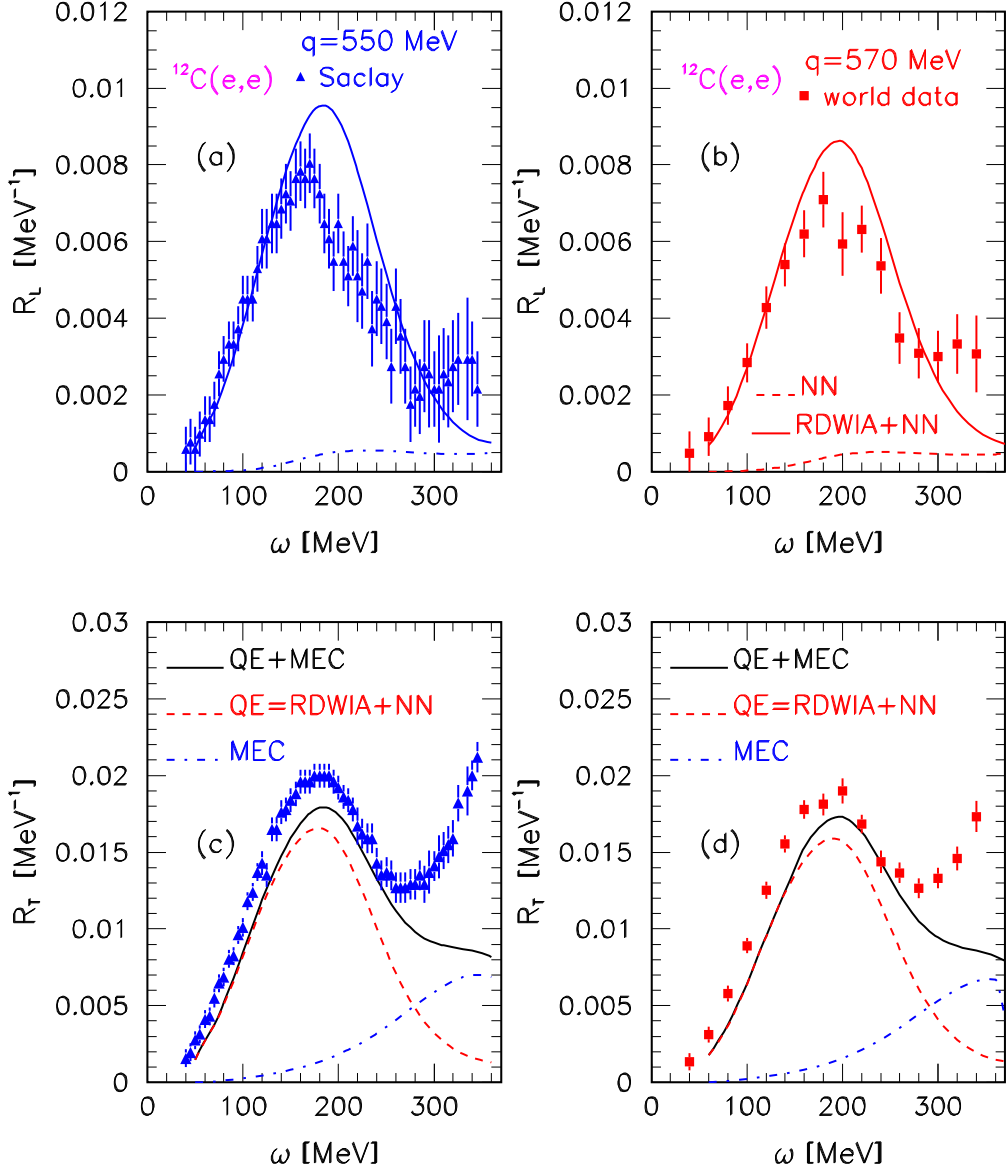}
  \end{center}
  \caption{\label{Fig3}(Color online) Same as in Fig.~\ref{Fig2}, but at 
    $|\q|=550$ MeV/c and $|\q|=570$ MeV/c}
\end{figure*}

The longitudinal and transverse response functions on carbon calculated in the 
RDWIA+MEC approach are shown in Figures~\ref{Fig1}-\ref{Fig3} for different 
values of the momentum transfer, together with Saclay~\cite{Saclay1} and the 
world data~\cite{Jour}. Note, that there are some differences between the two 
data sets because the world data exploited a wider range of the virtual photon 
polarization $\epsilon=0.05 \div 0.95$ for all $|\q|$-sets to reduce 
systematic errors in the Rothenbluth separation procedure. Also shows in 
figures are the contributions to $R_L(|\q|,\omega)$ from the $NN$-
correlated pairs and the contributions to $R_T(|\q|,\omega)$ from the 
$2p-2h$ MEC. It is worth niting that, the influence of the short-range 
correlations on the transverse response is considerably smaller than on the 
longitudinal one, because the $R_L$ is sensitive to the $NN$-correlations due 
to $NN$-interactions~\cite{Schi1, Schi2}.

The $R_L(|\q|,\omega)$ and $R_T(|\q|,\omega)$ responses as functions of energy 
transfer at $|\q|=300$ MeV/c are displayed in~Fig.~\ref{Fig1} with 
experimental results from Refs.~\cite{Saclay1,Jour}. Apparently, the 
calculation overestimates the value of the $R_L(|\q|,\omega)$ function, while 
the result for the $R_T(|\q|,\omega)$ response is in good agreement with the 
data. 
The longitudinal and transverse responses for $|\q|=400, 380, 500, and 570$ 
MeV/c are presented in Figs.~\ref{Fig2}, \ref{Fig3} and compared with data. The 
agreement between the RDWIA+MEC predictions and the world data is quite 
satisfactory. It is obvious that the inclusion of the $2p-2h$ MEC contributions
 increases the transverse responses at the high energy transfer and thereby 
improves the agreement with the data. 

\begin{figure*}
  \begin{center}
    \includegraphics[height=16cm,width=16cm]{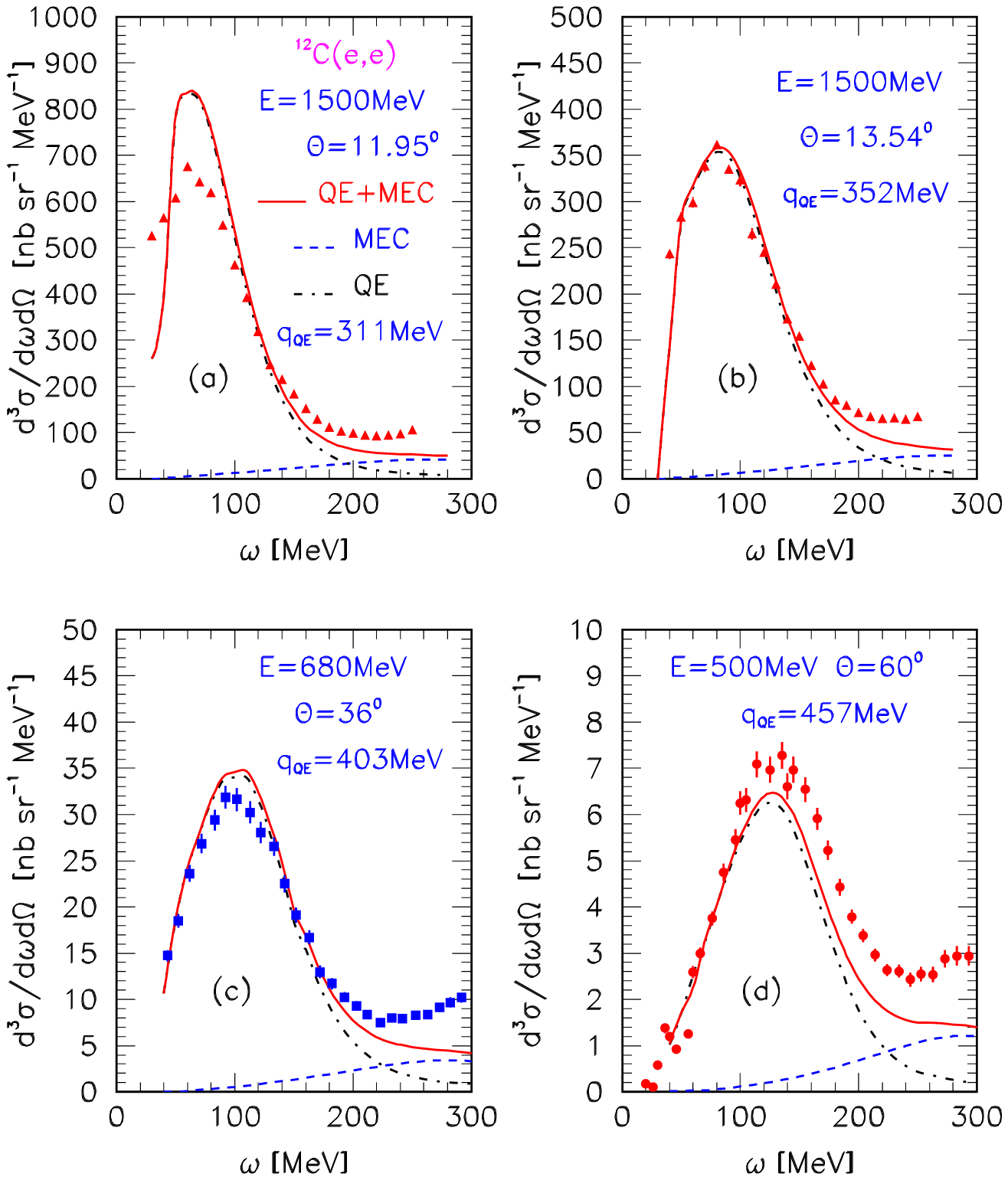}
  \end{center}
  \caption{\label{Fig4}(Color online) Inclusive cross section versus energy 
    transfer $\omega$ for electron scattering on ${}^{12}$C. The solid line is 
 the RDWIA + MEC results, the dashed line is the $2p-2h$ MEC contributions,
 and the dashed-dotted line is the contribution from the RDWIA . The data are 
from Ref.~\cite{Baran} (filled triangles), Ref.~\cite{Saclay1} 
(filled squares), 
    Ref.~\cite{Whit} (filled circles). In Ref.~\cite{Baran} data are for the 
    electron beam energy $E=1500$ MeV, and scattering angle 
    $\theta=11.95^{\circ},\theta=13.54^{\circ}$; in Ref.~\cite{Saclay1} data are 
    for $E=680$ MeV and $\theta=36^{\circ}$; in Ref.~\cite{Whit} data 
    are for $E=500$ MeV and $\theta=60^{\circ}$.}   
\end{figure*}

\begin{figure*}
  \begin{center}
    \includegraphics[height=16cm,width=16cm]{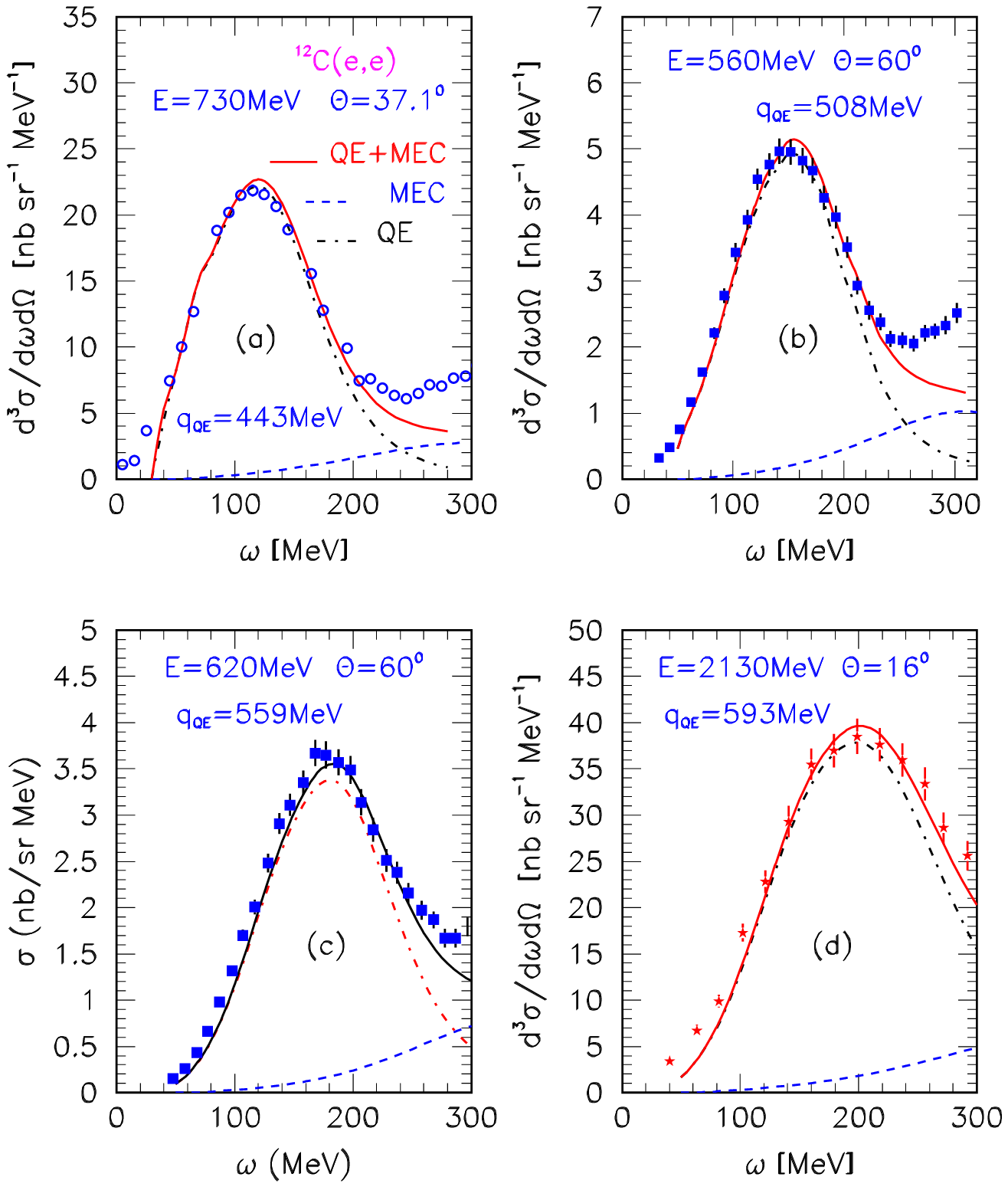}
  \end{center}
  \caption{\label{Fig5}(Color online) Same as in Fig.~\ref{Fig4}, but the data 
    are from Ref.~\cite{Ocon} (open circle) for the electron beam energy 
$E=730$ MeV and $\theta=37.1^{\circ}$; Ref.~\cite{Saclay1} (filled squares) for 
$E=560$ MeV and $\theta=60^{\circ}$ and $E=620$ MeV and $\theta=60^{\circ}$; 
Ref.~\cite{Ben1, Ben2} for $E=2130$ MeV and $\theta=16^{\circ}$.} 
\end{figure*}

To test the RDWIA+MEC approach we calculated the double differential inclusive 
${}^{12}$C$(e,e')$ cross sections versus the energy transfer to the nucleus. 
Results are shown in Figs.~\ref{Fig4}, \ref{Fig5} and compared with the data 
from Ref.~\cite{Whit,Ocon, Baran, Saclay1, Ben1, Ben2}. In each panel we show 
the contributions to the
inclusive cross section from QE and $2p-2h$ MEC processes. The comparisons 
were carried out for a wide range of kinematic variables, and each panel 
corresponds to fixed values of the incident electron energy and the scattering 
angle.
The panels have been ordered according to the corresponding value 
for the momentum transfer at the quasi-elastic peak $q_{QE}$. 
There is a good agreement between cross sections calculated in the RDWIA+MEC 
approach and experimental data, thus validating the 
reliability of our predictions. The positions, widths, and heights of the QE 
peaks are reproduced by this model. Only at particular kinematics, i.e. 
$E=1500$ MeV, $\theta=$11.95${}^{\circ}$, and $q_{QE}=311$ MeV/c~\cite{Baran} the 
calculation overestimates the value of the cross section and the 
underestimation of the data at QE peak occurs at $E=500$ MeV, 
$\theta=$60${}^{\circ}$, and $q_{QE}=$457 MeV/c~\cite{Whit} , while a good 
agreement is observed at clouse value $q_{QE}=$443 MeV/c, but at $E=730$ MeV 
and $\theta=$37.1${}^{\circ}$~\cite{Ocon}.  

It should be pointed out that the contribution of the $2p-2h$ MEC increases
 with the energy transfer and reaches its maximum in the ``dip'' region between 
the QE and the $\Delta$ peaks.
In these calculations we do not consider the inelastic contributions that can 
have an effect on the $(e,e')$ cross sections even in the QE regime. The 
inelastic part of the cross section is dominanted by the delta peak (mainly 
transverse) that contributes to the transverse response function. In 
particular, $\omega_{QE}=\sqrt{|\q|^2+m^2}-m$ corresponds roughly to the center 
of the quasielastic peak, $\omega_{\Delta}=\sqrt{|\q|^2+m^2_{\Delta}}-m$ to the 
$\Delta$-resonance [$m_{\Delta}$ is the mass of $\Delta(1232)$], and region 
between the two peaks to the two-body excitations. When the momentum transfer
 is not too high these regions are clearly separated in data
\begin{equation}
\Delta \omega=\omega_{\Delta}-\omega_{QE}= \frac{(m^2_{\Delta}-m^2)}
{\sqrt{|\q^2|+m^2} + \sqrt{|\q^2|+m^2_{\Delta}}},                   
\end{equation}
allowing for a test of theoretical models for each specific process.
At high momentum transfer the delta and QE peaks tend to overlap: in this 
case only the comparison with a complete model including inelastic processes is 
meaningful. In the present calculations only the real part of the Delta 
propagator is used and therefore the MEC peak is located in the range of  
$W\approx 1.14\div1.16$ GeV. However, in the ``frozen'' MEC approximation~
\cite{Amaro} with the full Delta propagator, (with real and imaginary parts) 
the $2p-2h$ MEC peak position is located near the $\Delta$ peak.

\subsection{Neutrino cross sections}

As was shown in Refs.~\cite{Martini1, Martini2, Nieves1, Nieves2, Megias2, 
BAV2, BAV5, MiniB1, MiniB2} two approaches can describe the enhanced cross 
sections observed in the MiniBooNE~\cite{MiniB1, MiniB2} CCQE data: 
one which 
includes an enhanced transverse response due to the $2p-2h$ MEC~\cite{Martini1, 
Martini2, Nieves1, Nieves2, Megias2} with $M_A \approx$ 1.03 GeV, and another 
is the impulse approximation approach ~\cite{BAV2, BAV5, MiniB1, MiniB2} with 
large value of  $M_A \approx$ 1.35 GeV. For incoming neutrino energy 
$\varepsilon_{\nu}=2$ GeV we calculated neutrino and antineutrino cross 
sections $(d\sigma/dx)_{QE+MEC}$ within the RDWIA+MEC model with $M_A=1.03$ GeV 
and $(d\sigma/dx)_{M_A,QE}$ cross sections in the RDWIA approach with $M_A=1.35$ 
GeV as functions of $x$, where $x=\varepsilon_{\mu}, Q^2, W$ are kinematic 
variables. To compare these distributions with the genuine CCQE 
$(d\sigma/dx)_{QE}$ cross sections, obtained in the RDWIA model with 
$M_A=1.03$ GeV we also calculated 
$R(MEC)=(d\sigma/dx)_{QE+MEC}/(d\sigma/dx)_{QE}$ and $R(M_A=1.35)=
(d\sigma/dx)_{M_A,QE}/(d\sigma/dx)_{QE}$ ratios.  

\begin{figure*}
  \begin{center}
    \includegraphics[height=16cm,width=16cm]{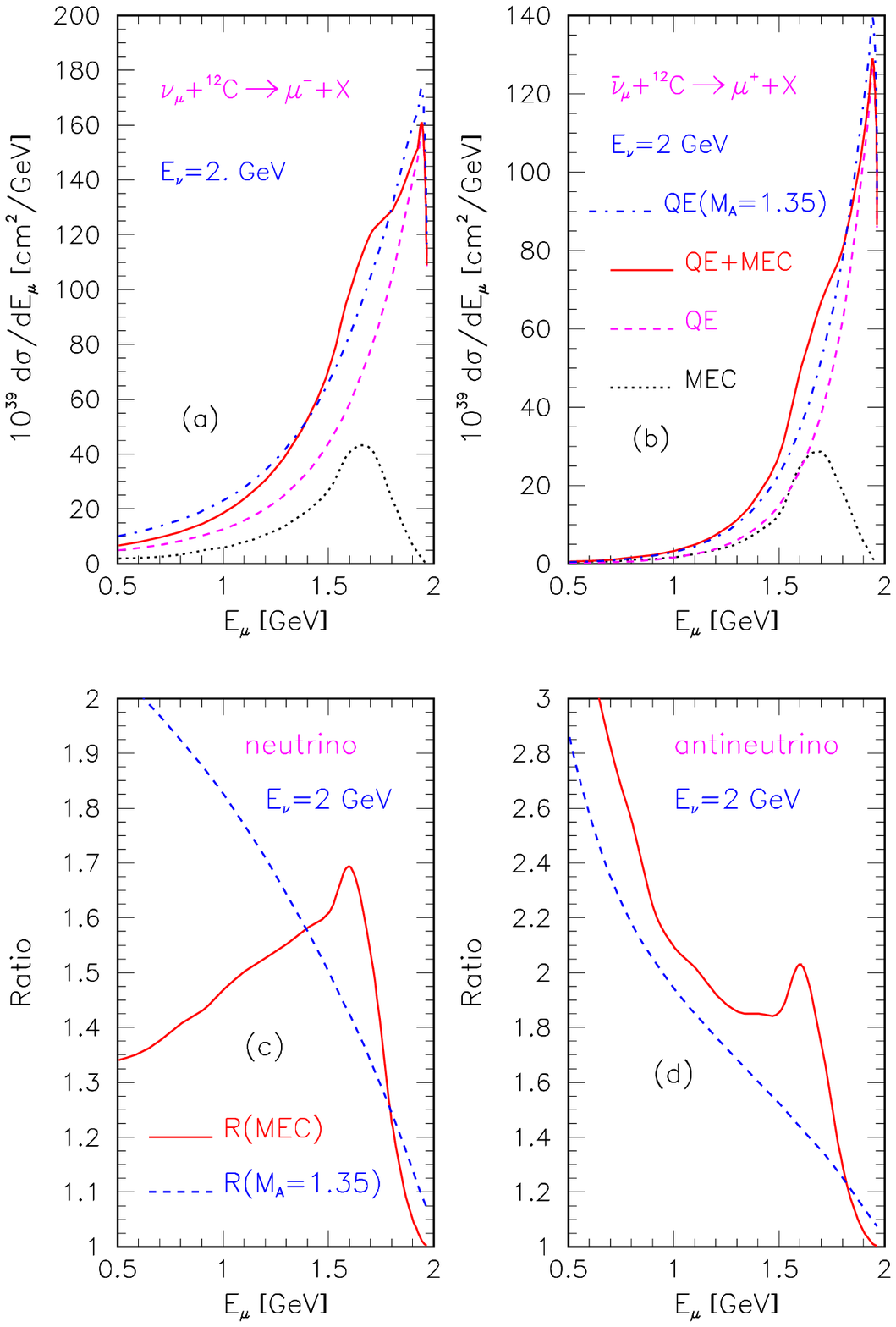}
  \end{center}
  \caption{\label{Fig6}(Color online) Inclusive cross section (upper panels) 
and ratios $R(MEC)$ and $R(M_A=1.35)$ (lower panels) vs the muon energy for 
neutrino and antineutrino scattering on ${}^{12}$C and for incoming neutrino 
energy $\varepsilon_{\nu}=2$ GeV. In the upper panels the solid line is the 
RDWIA+MEC calculation, the dash-dotted line is the RDWIA ($M_A=1.35$ GeV) 
calculation, whereas the dashed and dotted lines are the RDWIA($M_A=1.03$ GeV)
 and MEC contributions to the RDWIA+MEC cross sections. In the 
lower panel the solid and dashed lines are the $R(MEC)$ and $R(M_A=1.35)$ 
ratios, respectively.} 
\end{figure*}

The inclusive $d\sigma/d\varepsilon_{\mu}$ cross sections for neutrino and 
antineutrino scattering on carbon are presented in Fig.~\ref{Fig6} as functions
 of muon energy. Here, on the upper panels the results obtained in the 
RDWIA+MEC approach are compared with $(d\sigma/d\varepsilon_{\mu})_{M_A,QE}$ 
inclusive cross sections . Also shown are the contributions of 
the $2p-2h$ MEC and genuine CCQE process to the 
$(d\sigma/d\varepsilon_{\mu})_{QE+MEC}$ cross section. The 
lower panels show the $R(QE+MEC)$ and $R(M_A=1.35)$ ratios as functions of 
$\varepsilon_{\mu}$. One can observe that the $2p-2h$ MEC contribution increases 
with muon energy, reaching its maximum at $\varepsilon_{\mu}\approx$1.6 GeV, and 
becomes negligible in the region of the quasielastic peak. This leads to 
appearance of the peaks in the $R(MEC)$ ratios in the energy range 
$\varepsilon_{\mu}\approx 1.4\div 1.7$ GeV. Both models predict an increase of  
cross sections relative to the $(d\sigma/d\varepsilon)_{QE}$ results at low 
muon energies and show similar features near QE peak. Note that within the 
RDWIA model with $M_A=1.35$ GeV the cross sections in the region of the QE 
peak are predicted to be on $\approx 10{\%}$ higher than 
$(d\sigma/d\varepsilon)_{QE}$.

\begin{figure*}
  \begin{center}
    \includegraphics[height=16cm,width=16cm]{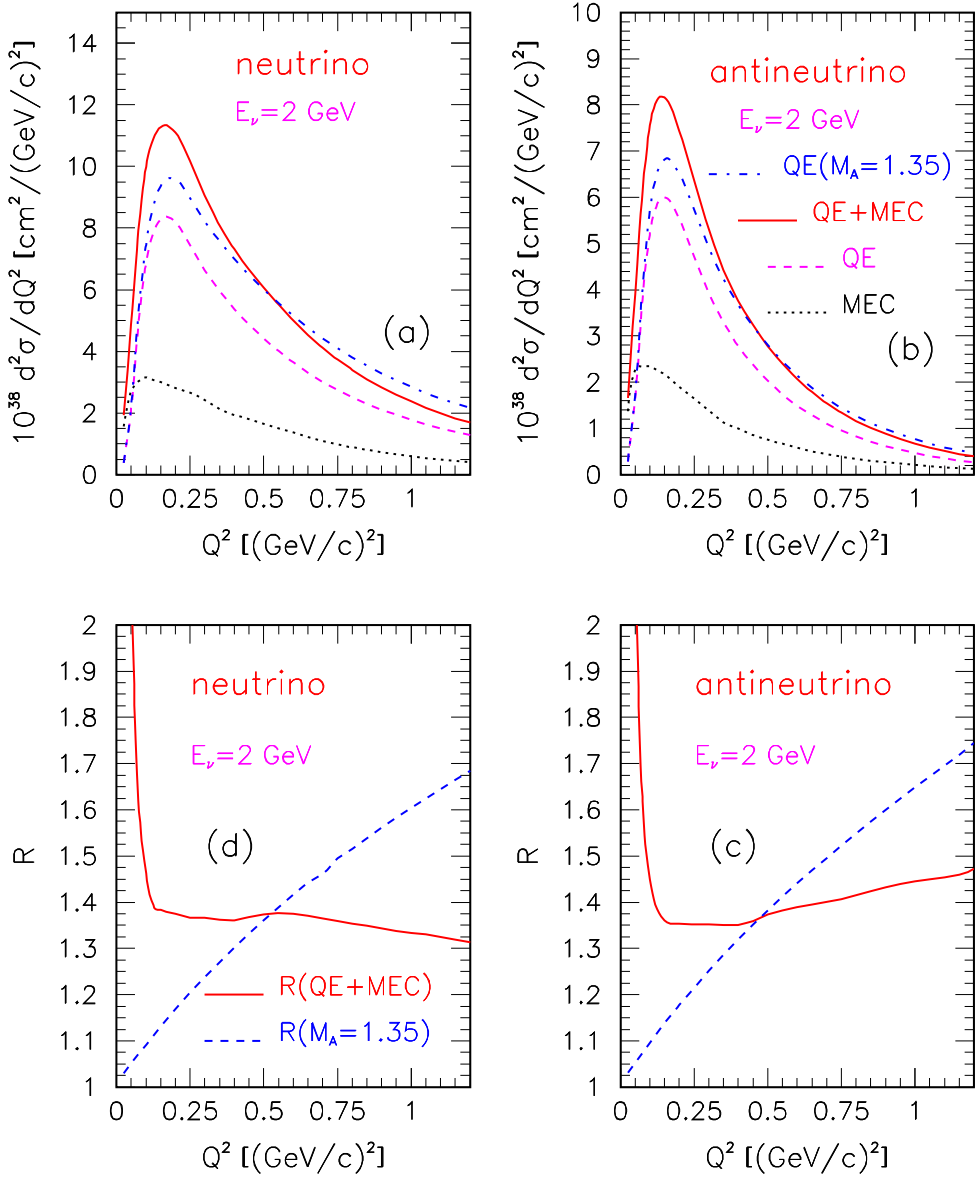}
  \end{center}
  \caption{\label{Fig7}(Color online) Same as in Fig.~\ref{Fig6}, but for 
$d\sigma/dQ^2$ cross section vs the $Q^2$.} 
\end{figure*}

Figure~\ref{Fig7} shows the same as Fig.~\ref{Fig6} but for $d\sigma/dQ^2$ 
cross sections as functions of $Q^2$. At $Q^2 < 0.2$ (GeV/c)${}^2$ the 
RDWIA+MEC model results are about two times larger than $(d\sigma/dQ^2)_{QE}$ 
cross sections. In the range $0.2 < Q^2 < 1$ (GeV/c)${}^2$ the ratio 
$R(MEC) \approx 1.4$ and slowly decreases (increases) with $Q^2$ for neutrino 
(antineutrino) scattering. Thus, in this $Q^2$ range the $2p-2h$ MEC 
contribution changes slightly the slopes of the $Q^2$-distributions calculated 
within the RDWIA model with $M_A=1.03$ GeV/c$^2$, because in the 
parametrization of the axial form factor $G^5_A$~(\ref{G5}) the value of 
$M_{A\Delta}\approx M_A \approx 1$ GeV is used. On the other hand, the 
$R(M=1.35)$ ratios increase with $Q^2$ from $R\approx 1$ at $Q^2\approx 0.1$ 
(GeV/c)${}^2$ to 1.7 at $Q^2\approx 1$ (GeV/c)${}^2$.  

\begin{figure*}
  \begin{center}
    \includegraphics[height=16cm,width=16cm]{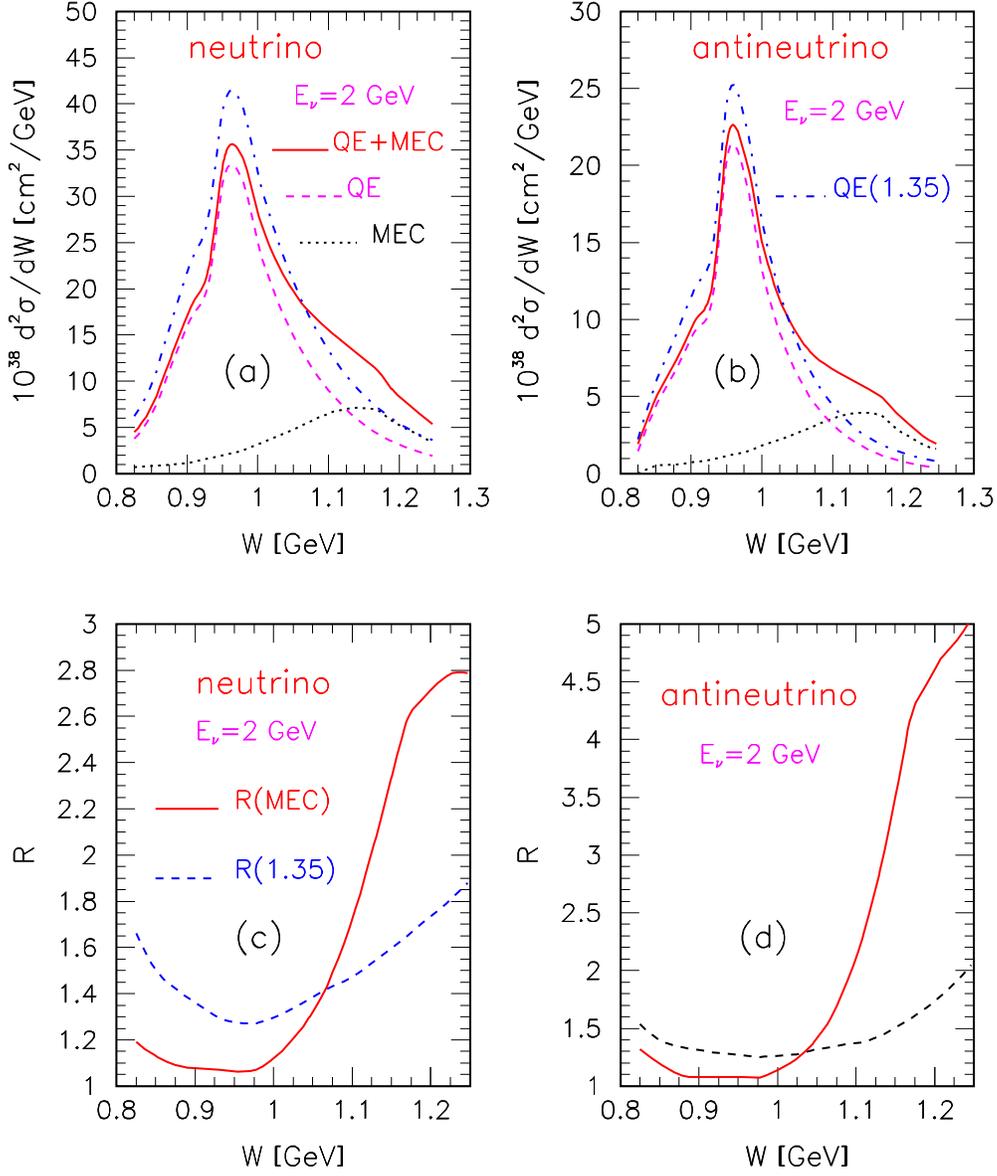}
  \end{center}
  \caption{\label{Fig8}(Color online) Same as in Fig.~\ref{Fig6}, but for 
$d\sigma/dW$ cross section vs the W.} 
\end{figure*}

Figure~\ref{Fig8} shows the same as Fig.~\ref{Fig6}, but for $d\sigma/dW$ cross 
sections as functions of W. The $2p-2h$ MEC contribution increases with 
invariant mass, and its maximum is located at $W \approx 1.15$ GeV, as in the 
case of electron scattering. The ratio $R(MEC)$ also increases with $W$ from 
$R\approx 1.1$ in the region of the QE peak up to 2.6(4.5) at $W=1.15$ GeV 
for neutrino (antineutrino) scattering. 
\begin{figure*}
  \begin{center}
    \includegraphics[height=16cm,width=16cm]{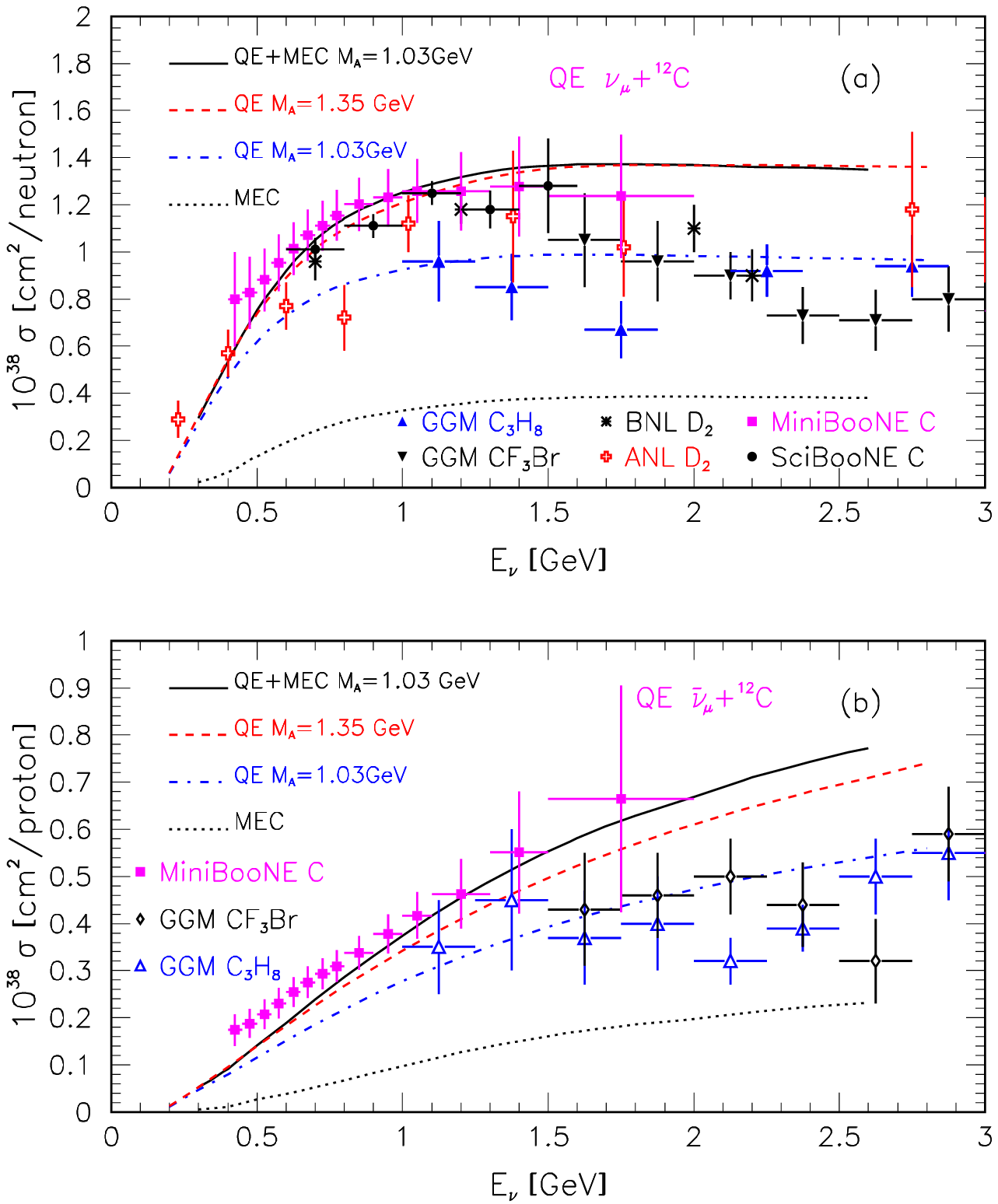}
  \end{center}
  \caption{\label{Fig9}(Color online) Total cross sections for QE and QE+MEC 
scattering of muon neutrino (upper panel) and antineutrino (lower panel) on 
${}^{12}$C as a function of incoming (anti)neutrino energy. Data points for 
different targets are from Refs.~\cite{MiniB1, MiniB2, SciB, Mann, Baker, 
Pohl, Brunner}. Also shown are predictions of the RDWIA+MEC (solid line),
RDWIA($M_A=1.35$ GeV) (dashed line), RDWIA($M_A=1.03$ GeV) (dash-dotted line),
 and $2p-2h$ MEC (dotted line).} 
\end{figure*}
At $W=0.94$ GeV the ratio $R(M_A=1.35)$ is $\approx 1.3$, and its slowly 
increases up to $\sim 1.6$ at $W\approx 1.15$ GeV. Apparently, the MEC-effects 
dominante in the ``dip'' region. 
 
The neutrino and antineutrino total cross sections $\sigma_{tot}$ together with 
data~\cite{MiniB1, MiniB2, SciB, Mann, Baker, Pohl, Brunner} are presented in 
Fig.~\ref{Fig9} as functions of the incoming neutrino energy. Here, the results
 obtained in the RDWIA+MEC approach are compared with the total cross sections 
calculated in the RDWIA model with $M_A=1.35$ GeV. Also shown are the RDWIA 
results with $M_A=1.03$ GeV, as well as contributions of the $2p-2h$ MEC that 
are about 27\% at $\varepsilon_{\nu} > 1$ GeV. The total cross sections are 
scaled with the number of neutron/proton in the target. From  
comparison of the RDWIA+MEC and RDWIA with $M_A=1.35$ GeV results it follows 
that the neutrino cross sections are in a good agreement and the difference 
between antineutrino cross sections is less than 
10\% at $\varepsilon_{\nu} > 1$ GeV.  

Thus, the analysis of the inclusive and total cross sections shows that 
the enhancement in either the transverse response, or in nucleon axial mass 
has almost the same effect on $d\sigma/d\var_{\mu}$ and total cross sections, 
and they are different for $Q^2$ and $W$-distributions.

However, $Q^2$ and 
$W$-distributions are not functions of direct observables, because $Q^2$ and 
$W$ are inferred kinematic variables which depend on incoming neutrino 
energy that is not known in the neutrino experiments with their broad 
incoming neutrino energy distribution. 
Most notably, neutrino energy reconstruction is possible only in model-dependent
 ways. Therefore, there is a growing interest in measurements of the hadronic 
system kinematics which allows one to increase the accuracy of the 
calorimetrical measurement of the incoming neutrino energy. All these 
developments will reduce our dependence on theoretical models.    
 
\section{Conclusions}

In this article, we studied quasielastic and $2p-2h$ MEC electron and 
(anti)neutrino scattering on a carbon target in the RDWIA+MEC and RDWIA with 
$M_A=1.35$ GeV approaches, placing particular emphasis on model dependence of
the inclusive $d\sigma/d\varepsilon_{\mu}$, $d\sigma/dQ^2$, $d\sigma/dW$, and 
total cross sections.

In the RDWIA+MEC approach we calculated quasielastic contributions to lepton 
scattering cross sections, using the RDWIA model with $M_A=1.03$ 
GeV and MEC electroweak response functions obtained in the RFGM. In calculation
 of the inclusive and total cross sections within the RDWIA, the effects of FSI 
and short-range $NN$-correlations in the target ground state were taken into 
account. An accurate parametrization of the exact MEC calculations of the 
nuclear response functions was used to evaluate the MEC response.
The RDWIA+MEC approach has been validated in the vector sector by describing 
the longitudinal and transverse response functions, as well as a set of 
inclusive electron scattering ${}^{12}$C data.  

We compared the inclusive cross sections for neutrino energy 
$\varepsilon_{\nu}=2$ GeV and total (anti)neutrino cross sections obtained in 
these approaches and found that while the enhancement in the transverse 
response or in axial mass have almost the same effects on inclusive 
$d\sigma/d\varepsilon_{\mu}$ and total cross sections this is not the case for 
the $Q^2$ and $W$-distributions, where two effects can be distinguished. On the
 other hand, $Q^2$ and $W$ are inferred variables that depend on neutrino 
energy. Therefore, one needs the new experimental approaches that used hadronic 
information in order to increase the accuracy of the calorimetric method of 
neutrino energy reconstruction in model-independent ways. 

\section*{Acknowledgments}

The authors greatly acknowledgs J. Amaro and G. Megias, for fruitful 
discussions  and for putting in our disposal the codes for calculation of the 
MEC's electroweak response functions that were used in this work. We specially 
thank A. Lidvansky and J. Samoilova for a critical reading of the manuscript.

%


\end{document}